\documentclass[letterpaper,12pt]{article}
\usepackage{graphicx}
\usepackage {amsmath}
\graphicspath{ {images/} }

\title{Patronage and power in rural India: a study based on interaction networks\thanks{This project is funded by the European Union under the 7th Research Framework Programme (Theme SSH.2011.1), Grant acknowledgement number 290752. Kar acknowledges additional financial and infrastructural support from Centre for Development Economics, Delhi School of Economics.} \thanks{We are indebted to the comments and suggestions of several seminar and conference participants.
Peter Lynn provided very useful help in our sample design. We acknowledge excellent research assistance from Vinayak Iyer, Amit Kumar, Mamta and Nishtha Sharma. Pramod Dubey and Pradeep Sahoo diligently monitored the household surveys. Of course, the errors are ours.} }

\author{Anindya Bhattacharya \thanks {Department of Economics and Related Studies, University of York.} ~~Anirban Kar \thanks{Delhi School of Economics, University of Delhi, Delhi, India.}~~Sunil Kumar \thanks{Freelance journalist and documentary maker, Delhi, India.}~ ~Alita Nandi~\thanks{ISER, University of Essex.} }

\begin{document}
\maketitle

\vspace{0.2in}
\begin{center}

{\large{[This version was actually produced on March 03, 2019.]}}

\end{center}

\vspace{0.2in}

\begin{abstract}
 
This work has two intertwined components: first, as part of a research programme it introduces a new methodology for identifying $\lq$power-centres' in rural societies of developing countries {\em in general} and then applies that in the {\em specific} context of contemporary rural India for addressing some debates on the dynamics of power in rural India. Land-ownership, caste hierarchy and patron-client relation have been regarded as the traditional building blocks of political-economic organization in rural India. However, many believe that gradual urbanization and expansion of market economy have eroded the influence of the traditional power structure. This work is a contribution toward identifying the nature of \lq local' rural institutions based on primary data collected by ourselves. We took 36 villages in the states of Maharashtra, Odisha and Uttar Pradesh - 12 in each of these states - as the sites for our observation and data collection. We quantify nature of institutions from data on the day-to-day interactions of households in the spheres of economy, society and politics. The aspect of institution we focus on is the structure of multidimensional and interlinked dependence in these spheres and whether such dependence is concentrated on a few \lq powerful' entities (called \lq local elites') dominating over a large number of households or whether this is distributed in a sufficiently diffuse manner. Our household survey shows that there is substantial variation in power structure across regions. We identified the presence of \lq local elites' in 22 villages out of 36 surveyed. We conducted a follow-up survey, called \lq elite survey', to get detailed information about the identified elite households. Our primary objective was to learn the socio-economic-political profile of the elite households and their involvement in village life. This paper provides a summary of our findings. We observe that landlordism has considerably weakened, land has ceased to be the sole source of power and new power-centres have emerged. Despite these changes, caste, landownership and patron-client relation continue to be three important pillars of rural power structure.

\end{abstract}

\vspace{0.2in}

\noindent {\bf JEL Classification:} O12, P16.

\vspace{0.2in}

\noindent {\bf Keywords and Phrases:} Clientelism, Lanlordism, Network

\newpage

\section{Introduction}

The motivation behind this work has two intertwined components: first, introducing a new methodology for identifying $\lq$power-centres' in rural societies of developing countries {\em in general}{\footnote{This work is one of the outputs of a research programme undertaken mainly by three of the authors, some other related outputs being Bhattacharya, Kar and Nandi (2017, 2018).}} and then applying that in the {\em specific} context of contemporary rural India for addressing some debates on the dynamics of power in rural India.  

There is a relatively recent surge of research in understanding clientelistic rural institutions: especially in the context of {\em political} clientelism. Bardhan and Mookherjee (2017) and Ravanilla et al. (2017) provide recent surveys of this strand of research. A particular theme of such research consists of identifying the position of political elites in some relevant social networks and then measuring the outcomes of having such a position (the study by Cruz et al., 2017 on Philippines is a notable example). However, in such studies, the clientelism-provider--the $\lq$elites'--are identified a priori. We, in contrast, take the network of socio-economic interactions (within and surrounding a village) as the {\em primitive} and then identify elites on the very basis of such network data (our methodology is described and its significance discussed in detail in Section 2 below). 

We apply this methodology to investigate a set of contentious issues in the political economy of rural India. Economic Survey of India (2017) estimates that about 377 million people from India's total population of 1.21 billion are city-dwellers. If urban-like features, high population density in a contiguous area and high density of build-up area, are used then the estimate is even higher. According to World Bank's agglomeration index, 55.3\%  of India's population already live in areas which have urban-like features.  Moreover, about 10 million people, as per Economic Survey of India (2017), migrate to cities and towns every year. There is a common perception among policymakers and academics alike that India is about to become, if not already is, a predominantly urbanized nation. Naturally, such features have given rise to the question of the nature of concomitant transformation of rural India.

Gupta (2005) asserts that \lq \lq the village in India, where life was once portrayed as \lq unchanging' and \lq idyllic', has in recent decades seen profound changes. The twin shackles that once decided matters for India's villagers, caste and agriculture, no longer exercise their vigorous hold." What are the channels of such transformation? Existing literature on this can be classified into three broad (but indeed interconnected) categories - economic changes, social changes and political changes. On the economic front two distinct channels have received particular attention - agrarian stagnation and gradual urbanization. So far as landholding is concerned, big landlord estates have mostly disappeared, average size of landholding has declined by 60 per cent in the last 50 years. The rural landscape now is believed to be characterized by cultivator-owners and landless labourers. In the meantime, profitability of farms, once boosted by technological changes during the \lq green revolution', has diminished to such an extent that by one estimate only families whose landholding are above four times of national average are able to earn a positive profit from cultivation (Basole and Basu, 2011)). Thus, agricultural employment both in the forms of casual labour and tenancy, once the most important sources of income for landless households, has become secondary to rural non-farm activities (like small business or self-employment) and urban casual employments. A CSDS study (2013) on the state of farmers revealed that 75 percent of farmers, particularly the younger generation, want to give up farming. Agricultural stagnation, along with a shift towards urban employment is understood to have triggered deep socio-political changes in rural areas. Earlier, landless households, particularly dalit and some OBC (other backward classes) castes, were tied to and dependent on the landed families for employment and credit. This was the main source of political power and social dominance of the landed castes. As agricultural stagnation pushed labouring castes outside the villages and out of agriculture, they have ceased to be dependent on the upper castes. Hence it is conjectured that the power of landlords as patron and political leaders has also diminished. Gupta (2005) claims that \lq \lq  caste as a system is dying" in rural India. Decline in power based on land and caste appears to have opened the door for a new type of political leaders. Krishna (2003) records that these \lq naya netas' (new leaders) lack inherited clientelistic networks. Instead. this is a group of educated youngsters whose main skill is the capacity to mediate between the administration and villagers. Jodhka (2014) asserts that one can observe the process of \lq dissociation, distancing and autonomy' in the emerging rural society: that is, clientelistic links based on an agricultural economy are disintegrating. Contest for power seems to have become more competitive and is being shaped by \lq macro' factors rather than local constraints.

Although several recent academic studies revolve around the above narrative, a debate has got stimulated too. Pattenden (2011) argues that although traditional route to power has weakened, more subtle kinds of political control by the upper castes persists. Ramachandran et. al. (2010) find persistence of landed power and wealth in their study. Harriss (2013) cautions that \lq \lq the idea that land is no longer so important as the basis of status and power should certainly not be overestimated." Ananthpur et al. (2014) observe that traditionally dominant households, even without formal positions in political office, can still influence decisions of local governments. Anderson et al. (2015 a,b) also document the presence and impact of clientelistic power relatiosn in the agrarian economy (of Maharashtra) through control over economic resources.

In this backdrop, our study aims to address systematically the following questions in the context of India. Has clientelistic power relations, embedded in an agrarian economy, lost its prominence in rural India or does it still survive in a more subtle and diversified form? Are there \lq local' powerful elements who control patronage networks? If so, are they mainly the erstwhile landlords or are they predominantly new entrants? What is the source of their power: is it mainly their ability to mediate between administration and villagers or is it their control over important economic resources? Our approach in addressing these questions brings together aspects of qualitative village studies and quantitative economic modelling. We believe that institution is not {\em merely an economic} category - it is an aggregate of social relations that operate within political-economic spheres. Hence nature of rural institutions cannot be identified by merely looking at economic indicators (as attempted in, for example, Basole and Basu, 2011). We consider it to be more illuminating to measure the nature of institutions from data on the personalized day-to-day interactions of agents in the spheres of economy, society and politics. We do so on the bases of primary data collected by ourselves in 36 villages from three states in India (details below).

In the next section we describe our methodology and data. Section 3 discusses the results of our analyses. Some concluding comments are given in Section 4. The relevant tables and diagrams are collected at the end. At the very end, in an appendix we describe our sampling methodology in detail.

\section{Methodology and Data}

\subsection{The conceptual framework}

We primarily conceptualize institution as the structure of multidimensional dependence in the spheres of economy, society and politics. For investigating the nature of rural power relations (and clientelism in particular), we seek to explore whether such dependence is concentrated on a few entities dominating over a good many of the households or whether this is distributed in a sufficiently diffuse manner. Moreover, our emphasis is on such dependence at {\em local} level, roughly at the level of a village and the neighbouring villages and town(s). Meiksins Wood (2003) emphasizes this aspect while distinguishing between institutions. She observes \lq\lq ...only capitalism has a distinct economic sphere. This is so both because economic power is separate from political or military force and because it is only in capitalism that the market has a force of its own, which imposes on everyone, capitalists as well as workers, certain impersonal systemic requirements of competition, accumulation and profit-maximization...Although the sovereign territorial state was not created by capitalism, the distinctively capitalist separation of the economic and the political has produced a more clearly defined and complete territorial sovereignty than was possible in non-capitalist societies. At the same time, many social functions that once fell within the scope of state administration or communal regulation now belong to the economy''. While we do not seek to label institutions as \lq capitalist' or \lq semi-feudal' etc, we use the pattern of dependence as an identifier of changing institutional dynamics.

The basis for constructing our institutional variation is the multidimensional directed network where the primary nodes are the sampled households (HH hereafter) in villages{\footnote{This section draws heavily from Bhattacharya, Kar and Nandi (2017) a companion paper of ours.}}. We collected data from 36 villages (selected through stratified and then PPS sampling) in three states of India (12 villages from each of Maharashtra, Odisha and the (eastern half) of Uttar Pradesh (UP)) using personal interviews at the HH level\footnote{A part of the HH survey, covering mainly Odisha and Maharashtra, took place during March-April, 2013 and much of the survey in UP took place in November-December, 2013.}. While we provide the details of our sample design in an appendix at the end, here we still make a few remarks about the representativeness of our sample. First, one of the states, Maharashtra, happens to be among the richest states in India (containing the city of Mumbai, the commercial capital of India) whereas the state of Odisha is considered to be a poor one: with histories of famine and starvation-deaths in some of its rural parts. Next, given the well-known evidence of `persistence' of land-revenue-related institutions from the British colonial period in rural India (e.g., Banerjee and Iyer, 2005), we took care so that our sample villages contain diverse histories of such land-revenue-related institutions: {\em ryotwari} (direct payment by the farmers to governmental tax-collectors) to {\em zamindari} (landlord-based system) as well as erstwhile princely states. Further, since the structure of rural economy in interior areas (which is expected to be predominantly agricultural) is quite different from that in coastal villages (where fishing and related activities generate a good deal of diversity of occupational structure), we included both coastal and interior villages in our sample. Therefore, while our sample size for villages is not much large, these are sufficiently representative for a large part of India.  

Since we wanted to capture the network of interactions in each village sampled, we adopted the following scheme: if a sample village contained upto 100 HHs, then all the HHs in that village were surveyed; and with sample villages with more than 100 HHs, approximately 100 randomly selected HHs were surveyed. In the HH-survey we gathered information, among other aspects, on links the sample households have for help in spheres of day-to-day economic interactions (like whom the HH depends on primarily for getting productive inputs, for selling of outputs if any, for loans etc), social interactions (like whom the HH approaches for advices on family matters and disputes, religious matters etc) as well as political ones (like whom the HH accompanies to political events if any etc). In Table1a we give the detailed list of the items of such help. We aggregate this information on multidimensional linkages of HHs on other HHs or entities to derive an unidimensional {\em dependence} network in each village (the details of which are given in the following sub-section). The main underlying principle for constructing these dependence networks is that a HH A is \lq dependent' on a HH B, when it receives certain economic/social/political input from another HH B and the converse is not true. An entity with sufficiently many such dependents is called a local {\em elite}. A village having at least one local elite is called an {\em elite village}{\footnote{In Bhattacharya, Kar and Nandi (2017) we also assigned a score of `elitism', called $Nclscore,$ to each village which measures the {\em intensity} of clientelistic institutions prevailing in that village. But we refrain from a description of that here as that is not strictly relevant for the present exercise.}}. Recall from Section 1 that most of the studies which try to identify power centres, either rely on {\em direct queries about important entities}, which is vulnerable to misreporting and under-reporting, or are based on the researchers' subjective impressions. We emphasize that in contrast, we systematically derive the set of \lq important' entities from the revealed behaviour of the respondents themselves in several spheres of their actual lived experiences.

One noticeable feature of what we call an elite village is that such a village consists of a small number of persons/households (in our data we find them to be usually less than four or five) who have control over a number of households in terms of crucial economic dimensions (like providing credit or employment) and very often these same persons dominate in the spheres of social interactions as well as in political arenas around the village. In many cases the village {\em sarpanch/pradhan} (head) happens to be {\em one} such person\footnote{Very similar institutional features are corroborated by Ananthpur et al. (2014) in their micro-study in Karnataka (a state different from the ones from which we collected data)}. However, as we note in detail in the following section, not all elites have connections with formal political power.

Note that local elites were identified on the bases of HH responses in our HH survey and they themselves may not have been part of the HH sample. Therefore we conducted a second round of survey to interview such elites/elite HHs personally \footnote{This took place one year after the HH survey--in the second half of 2014.}. This survey will be referred to as \lq Elite survey'. The core of the analysis of this paper, on the characteristics and behaviour of the local rural elites, is mainly based on this  Elite survey. 

\subsection{Localized institution: measuring dependence, elites, clients}

Our quantitative identification of the elites is based on the following core underlying principles. First, dependence is embedded in day to day activities, both economic and socio-political. Access to inputs of production, market access for products, dispute resolution and participation in political process are a few examples of such activities. The second aspect of dependence is {\em personalized interaction}. This is distinct from formal institutional interactions. Borrowing from banks, approaching police stations for dispute resolution etc. are instances of formal institutional interactions, while borrowing from informal lenders, approaching local political leaders for dispute resolution are personalized interactions. These two aspects together imply that the dependence structure we are exploring is essentially localized in nature. Finally, high concentration and interlinkage of dependence links are indicators of stronger localized power.

Recall that the primitive in this context in our set-up are the households' links for getting help in social, economic and political spheres. If HH $M$ receives an economic, social or political service from HH $N$, then HH $M$ is said to have an {\bf outgoing service link} to household $N$. We also classify outgoing service links into two groups - {\em{crucial and non-crucial}} - based on their relative importance. This classification is based on our perception and judgment. For instance, a service-link of seeking advice for resolution of household disputes is categorized as non-crucial whereas seeking advice for profession-related disputes is categorized as crucial. Admittedly, this classification is subjective but not arbitrary. The full list of services, classified as $(i)$ economic/social/political and $(ii)$ crucial/non-crucial is given in Table 1a. Since we asked each surveyed household whether such services are reciprocated, we also have data on outgoing links from Household $N$ to Household $M$. In case, Household $M$ is also part of our sample, we have an independent verification of such claims (we could not make such cross-verification {\em in general} though)\footnote{In case of mismatch, though such instances are rare, claims of the household which has received the service is accepted.}.

Note that, in our network data, there can be multiple such service-links between two nodes: i.e., households. First we aggregate these to a single dimension, called {\bf dependence-connection}. To capture relative strength of dependence relation, we classify dependence-connections into three types.

\noindent
Type $A$: HH $M$ is said to have Type $A$ outgoing dependence-connection to HH $N$ only if $M$ has exactly one crucial outgoing service-link to $N$. A single non-crucial link is unlikely to be an indicator of clientelistic relation.\\
Type $B$: HH $M$ is said to have Type $B$ outgoing dependence-connection to HH $N$ only if $M$ has at least two outgoing links to $N$ that are of similar kind, either all economic or all social or all political. This captures interlinkages in received services .\\
Type $C$: HH $M$ is said to have Type $C$ outgoing dependence connection to HH $N$ only if $(i)$ $M$ has at least two outgoing links to $N$ and $(ii)$ not all of them are of similar kind (economic/social/political) of services. This captures interlinkage in different spheres of daily/usual interactions.

Clearly we make the definition of dependence more demanding as we move from Type A to Type C. Since (clientelistic) dependence should be conceived as an asymmetric power relation (in contrast to a reciprocal relationship like friendship), we exclude all bilateral, mutual outgoing dependence-connections from our network. Directed cycles of higher length are rare in our data. Thus HH $M$ is said to be {\bf dependent} on HH $N$ if $(i)$ HH $M$ has a dependence-connection of at least one type to HH $N$ and $(ii)$ HH $N$ does not have any type of dependence-connection to HH $M$.\\
This completes the description of {\em dependence network} for a village.

Next we use this derived network data to identify the presence and pattern of clientelist institution in surveyed villages. If a clientelist network is present then it would be characterized by patrons and clients. It is expected that clients will be dependent on patron(s) for various (often interlinked) services and a number of clients will be dependent on a patron. Thus a \lq hub and spoke' type network is expected to emerge in such villages. To this end, we define a patron, called {\bf elite} as follows. {\em If more than 5\% of the sampled households are dependent on a household X then X is potentially an important patronage-provider in the village and is called an elite}. This captures concentration of dependence in our network data. A household which is dependent on at least one elite is called a {\bf client}. Any household, which is neither a client nor an elite will be called a {\bf non-client}. In Figures 1 and 2 (given at the end) we show dependence-connection networks of two villages, one with presence of elites and another without.

A couple of comments are due at this stage. First, since we have not done census for all of the sample villages, we only see the village dependence networks partially. We can not rule out presence of additional hubs in the dependence network for some villages and our elite-identification could be incomplete. Second, as mentioned above, other than mutual dependence, higher order directed cycles are rare in our village networks. Therefore, more complicated network centrality measures (as in, e.g., Herings et al., 2005) are somewhat superfluous for our purpose and we choose to stick to the simple definition of elites described above.

In our sample, we observe considerable heterogeneity across regions. Out of 36 sampled villages, 14 villages have no elites and 22 villages have at least one elite. We have identified 50 elites and 480 clients in 22 villages. State-wise division of elites is as follows: 15 elites in 8 villages of (Eastern) Uttar Pradesh, 16 elites in 6 villages of Maharashtra and 19 elites in 8 villages of Odisha. During the \lq Elite survey', we tried to interview all the elites but due to refusal and non-availability, we could meet only 40 of them - 14 in (Eastern) Uttar Pradesh, 7 in Maharashtra and 19 in Odisha.

\section{Results}

\noindent Even before getting into the discussion about the characteristics and behaviour of the elite HHs, the first striking feature which emerges from the last paragraph is that {\em the presence of clientelistic institutions, as measured by us, is not uniform.} In each of the states both elite and non-elite villages exist! Thus, rural institutions in India do not seem to be prone for any sweeping generalizations.

\subsection{Profile of elite households: their economic decisions}

\noindent {\em{Elite and non-elite households are different in terms of some $\lq$endowment' characteristics}}\\

\noindent First we focus on a few endowment-like characteristics of elites and non-elite households. A summary of these differences is available in Table 3. We find that compared to non-elites, elites are more likely to belong to upper castes, have more land-holding, have higher non-land wealth score, more likely to be in stable occupations and more likely to have a royal or landlord heritage. These differences are statistically significant at 1\% level of significance (based on two sided t-tests also reported in Table 3).\\ 
Specifically, we find that while 82.5\% of elites are either of upper caste or OBC (the rest being SC/ST and Muslims), only 52.8\% of non-elites are so.\\ 
Occupational pattern of elite households reveals that more than 50\% of such households have a steady source of income (which is measured as the main occupation being a salaried position or operating a business or a factory). For the non-elite households, the corresponding fraction is drastically lower: only about 7\%.\\ 
The average size of land-holding in acres for elite households is 0.842{\footnote{However, informal anecdotal discussion within some villages revealed that land-holding of several elite HHs were under-reported as is often usual in Indian villages.}} but less than half that (0.348) for non-elite households.\\ 
We also construct the non-land assets wealth index by taking the sum of five dummy variables each representing whether the household has one type of non-land asset like automobiles, tractor etc (see the detailed definition in Table 2). Average non-land wealth-index score for elite household is 2.58 (out of a maximum of 5) compared to 1.01 for non-elite households.\\ 
These comparisons show that our identified elite households are mainly from wealthy and socially dominant sections. Moreover these features also indicate the asymmetric nature of the dependence-connections as a result of which our dependence network is quite different from other kinds of social networks (for example, like a friendship network) that have homophily as a primary feature.\\

\noindent {\em{Occupational structure in elite households and its dynamics}}\\

\noindent 81\% of elite households report that they still consider farming as one of their main sources of earning. However 30\% of such households report business or factory production as another main source of the household income.Only 3\% of the elite households depend primarily on sources other than farming or business (mostly salaried). Therefore, elite households have {\em diversified} quite a bit into non-agricultural occupations. This finding is in conformity with the recent literature. However, it is also clear that elite households have not moved away substantially from agriculture yet.\\
In Table 4 we split the distribution of occupation along age (below and above 40) and gender. One quite interesting finding is that among the young females within such HHs there is a considerable rise in acquisition of skills and/or taking up business-related occupation.\\    

\noindent {\em Landholding and its dynamics}\\

\noindent So far as land-ownership is concerned, we believe that there is huge underreporting in our data as well as in official statistics. It is instructive to note that India may have unique identification number for every citizen but its land record data is notoriously poor. As reported by the elite households, at least 84\% of the households have at least one acre of land. However, informal conversation in some villages in Maharashtra and UP (outside the formal interviews) throw up figures like between 30 to 50 acres for some households. Still, these numbers are nowhere close to that of erstwhile big landlords. Among the elite HHs only 30\% of them report that their ancestors  were erstwhile royals or landlords. So it is safe to conclude that land has ceased to be the {\em sole} source of power. But at the same time, local elites are holding on to their land. Below 10 percent of elite households admitted that they have sold land in last 10 years. In contrast, about 13\% of elite households reported that they purchased land in the last 5 years. Further, when land has been sold, never was the purpose some non-agricultural investment. The purposes were either for purchasing some other piece of land or for some temporary household need. Therefore, in spite of increasing diversification of occupations within such households, land remains a primary (but not the only) form of asset.\\

\noindent {\em Nature of investment decisions}\\

\noindent We asked each elite household some details of the nature of their investment decisions in the last five years. Quite startlingly, 62\% of such households do not report any investment at all for that time span. Among those who invested, 32\% of the households made investment for productive purposes like purchasing new tractors/harvesters, or starting some new business etc. Only 8\% of the households invested in purchasing stocks or shares or gold or in local lending business. Therefore, rural elites, it seems, in spite of some occupational diversifications, are still not substantially into entrepreneurial activities.\\
One noticeable feature is that both among the school-going and college-going members of such households, fraction of those attending (usually more expensive) private institutions is higher than those attending government institutions. This indicates that the elite households are investing into human capital for their children too.  

\subsection{Source of power?}

Krishna (2003) attributed the rise of new $\lq$elite'-entrants to decentralization of governance. He argued that educated youth found new political space which was hitherto unavailable. Their capacity to mediate between administration and villagers gave them electoral advantage over traditional elites. Our findings do not entirely support the above narrative. Only about 46$\%$ of elites have held panchayat (rural local government) positions or were political party functionaries. Thus it seems that political office is not a necessary condition for political influence. For a couple of villages in UP, anecdotal evidence in fact suggests that the $pradhan$ (head of the village-level local government) is actually a dummy for the local traditional landlord elements. But we found evidence about the opposite phenomenon too, especially in Maharashtra. About 15\% of the elites hold position in religious and caste organizations, which are informal bodies. Moreover, 50\% elite households have relatives in influential social and political positions, which indicates that dynastic power matters as much as individual entrepreneurship in the arena of formal politics.

Finally we try to locate the elites' sphere of influence as reported by the elites themselves. Elites typically have control over and are involved in disbursal of different kinds of resources. We classify such activities (the $\lq$local sevices' provided) into three categories: economic (E), political (P) and social (S). Details are available in Table 1b. We find that 61\% of the elite households are active in all of the three spheres: E, P and S (details about the pattern of patronage-service provision are summarized in Table 5). Quite strikingly we find that elite households providing only economic services constitute 18\% of our cohort. However, there exists no elite households which provide only political or social services. Thus, the notion that control over economic resources has become unimportant for understanding the contour of rural power is not borne out by our investigation. Rather, we find that control over economic resources, a variety of them - not just land, still plays a significant role. Moreover, exercising influence over different interlinked spheres of rural life is clearly important.

\section{Summary and concluding remarks}

The first point to note is that the rural elites identified via our methodology conform partially, but not totally, with the common subjective perceptions: that such HHs are predominantly wealthy, having stable sources of income, having often, but not always connection with formal political power etc. Thus, our work is like a counterpart to the interesting findings in the research by Cruz et al. (2017) (also notable in this context is Naidu et al., 2017). Secondly, in the context of India we find substantial variation in terms of the prevalence of clientelistic institutions within rural societies. Thirdly, the rural elites, power centres within the villages and surrounding localities, cease to be exclusively from the erstwhile landlord lineage, but still hold land as one of the primary forms of economic asset. However, in such HHs, quite a bit of occupational diversity has taken place, but that has failed to induce many such HHs into productive investments. Finally, the source of power of such HHs usually straddles several spheres of rural life and never in our study do we find that the source of power is exclusively in social or political spheres.

A natural question is of measuring the impact of such clientalistic institutional variations on outcomes. These are taken up, in the context of HH-level provision of workfare-jobs, in Bhattacharya, Kar and Nandi (2017) and in the context of village-level provision of mainly public sevices in Bhattacharya, Kar and Nandi (2018).\\

\section*{References}
\begin{enumerate}
\item Ananthpur, K., K. Malik and V. Rao (2014); The anatomy of failure: An ethnography of a randomized trial to deepen democracy in rural India; Mimeo, World Bank, Report No. WPS6958.
\item Anderson, S., P. Francois and A. Kotwal (2015a); Clientelism in Indian Villages; American Economic Review; Vol. 105, No. 6, 1780-1816.
\item Anderson, S., P. Francois, A. Kotwal and A. Kulkarni (2015b); Implementing MGNREGS: \lq One Kind of Democracy'; Economic and Political Weekly, Vol. 50, Issue 26-27, 44-48.
\item Banerjee, A. and L. Iyer (2005); History, Institutions, and Economic Performance: The Legacy of Colonial Land Tenure Systems in India; American Economic Review, Vol 95, No. 4, 1190-1213.
\item  Bardhan, P. and D. Mookherjee (2017); Clientelistic Politics and Economic Development: An Overview; Mimeo, http://people.bu.edu/dilipm/\\wkpap/EDIclientsurvMay17Fin.pdf
\item Basole, A. and D. Basu (2011); Relations of Production and Modes of Surplus Extraction in India: Part I - Agriculture; Economic and Political Weekly, Vol. 46, Issue 14, 41-58.
\item Bhattacharya, A., A. Kar and A. Nandi (2017); Local institutional structure and clientelistic access to employment: the case of MGNREGS in three states of India; ISER Working Paper No. 2017-14; https://www.iser.essex.ac.uk/research/publications/working-papers/iser/2017-14.pdf.
\item Bhattacharya, A., A. Kar and A. Nandi (2018); Does clientelism adversely affect rural Infrastructure? An empirical study; (in preparation).
\item Cruz, C., J. Labonne and P. Querubin (2017); Politician Family Networks and Electoral Outcomes: Evidence from the Philippines; American Economic Review; Vol. 107, No.10, 3006-37.
\item Economic Survey of India (2017); Economic Survey 2016-17. Government of India. Ministry of Finance. Department of Economic Affairs. Economic Division.
\item Gupta, D. (2005); Whither the Indian Village: Culture and Agriculture in \lq \lq Rural" India; Economic and Political Weekly, Vol.40, Issue 8, 751-758.
\item Harris, J. (2013); Does \lq Landlordism' Still Matter? Reflections on Agrarian Change in India; Journal of Agrarian Change, Vol. 13, Issue 3, 351-364.
\item  Herings, P.J., G. van der Laan and D. Talman (2005); The Positional Power of Nodes in Digraphs; Social Choice and Welfare, Vol. 24, 439-454.
\item Jodhka, S. (2014); Emergent Ruralities: Revisiting Village Life and Agrarian Change in Haryana; Economic and Political Weekly, Vol. 49, Issue 26-27, 5-17.
\item Krishna, A. (2003); What is Happening to Caste? A View from Some North Indian Villages; Journal of Asian Studies, Vol. 62, Issue 4, 1171-1194.
\item Lokniti (2013); State of Indian Farmers: A Report. Delhi. Available at www.lokniti.org/pdf/Farmers\_Report\_Final.pdf
\item Pattenden, J. (2011); Gatekeeping as Accumulation and Domination: Decentralisation and Class Relations in Rural South India; Journal of Agrarian Change, Vol. 11, Issue 2, 164-194.
\item Naidu, S., J.A. Robinson and L.E. Young (2017); Social Origins of Dictatorships: Elite Networks and Political Transitions in Haiti; Mimeo, http://www.laurenelyssayoung.com/wp-content/uploads/2016/11/\\NaiduRobinsonYoung\_Haiti.pdf
\item Ramachandran, V.K., V. Rawal and M. Swaminathan, eds (2010), Socio-Economic Surveys of Three Villages in Andhra Pradesh; New Delhi: Tulika Books.
\item Ravanilla, N., D. Haim and A. Hicken (2017); Brokers, Social Networks, Reciprocity, and Clientelism; Mimeo, http://dotanhaim.com/wp-content/\\uploads/2017/09/Ravanilla-et-al\_APSA2017-paper.pdf
\item Sbriccoli, T. (2016); Land, Labour and Power: A Malwa Village , 1954-2012; Economic and Political Weekly, Vol. 51, Issue 26-27, 8-16.
\item Wood, E. M. (2003); Empire of Capital; Verso.

\end{enumerate}

\newpage

\section*{Tables and Diagrams}

\vskip3em

\begin{tabular}{|p{6cm}|p{3cm}|p{2cm}|}
 \hline
 \multicolumn{3}{|c|}{Table 1a: Types of services (from the HH survey)} \\
 \hline
 Type of services & Economic, social or political & Crucial or non-crucial\\
 \hline
Lease in land or sharecropping & Economic & Crucial\\
Purchanse input of production & Economic & Not crucial\\
Sale output & Economic & Crucial\\
Getting employment & Economic & Crucial\\
Getting information on MGNREGS & Economic & Non-crucial\\
Paying bribe for governmental welfare services & Economic & Crucial\\
Assistance for welfare & Political & Crucial\\
Household related dispute mediation & Social & Not crucial\\
Employment related dispute mediation & Social & Crucial\\
Guidance on political matter (like whom to vote or accompanying to political meetings or rallies) & Political & Crucial\\
Guidance on religious matter & Social & Not crucial\\
\hline
\end{tabular}

\vskip1em

\begin{tabular}{|p{4cm}||p{8cm}|}
 \hline
 \multicolumn{2}{|c|}{Table 1b: Types of patronage services (from the Elite survey)} \\
 \hline
 Type of services & Details\\
 \hline
Economic services (E) & Rent out land for farming; \newline Rent out farm inputs; \newline Provides employment; \newline Lends money as an ancillary service; \newline Runs a rural business.\\
\hline
Political services (P) & Helps villagers with government schemes; \newline Offers politics-related advices; \newline Organizes political rallies; \newline Mediates between administration and villagers; \newline Mediates in occupation-related disputes between villagers.\\
\hline
Social services (S) & Mediates in family disputes; \newline Organizes religious/charity/cultural events; \newline Offers religion-related advices.\\
\hline
\end{tabular}

\newpage

\begin{tabular}{|p{4cm}||p{8cm}|}
\hline
\multicolumn{2}{|c|}{Table 2: Detailed definition of some variables} \\
\hline
Name of the variable & Definition\\
\hline
Landholding & Dummy; 0 if reported landholding \newline of the HH is less than 1 acre \newline and 1 otherwise \\
\hline
Wealth index & Sum of dummies; underlying \newline dummies take value 1 if \newline HH has a brick-built house \newline HH has a flat/house in town \newline HH has a palang (ornate bed) \newline HH has tractor(s)/harvester(s) \newline HH has an automobile\\
\hline
Upper caste & Dummy: takes value 1 if the \newline HH is Brahmin or general or OBC.\\
\hline
Stable occupation & Dummy: takes value \newline 1 if the HH \newline reports its main \newline occupation as business \newline or salaried profession\\
\hline
\end{tabular}

\newpage

\begin{tabular}{|p{3cm}||p{1cm}|p{1cm}|p{2cm}|p{2cm}|p{2cm}|  }
 \hline
 \multicolumn{6}{|c|}{Table 3: Elites and non-elites: difference in certain crucial characteristics} \\
 \hline
 Variable & Mean for elites & Mean for non-elites & Difference significant at what level & No. of observations (elites) & No. of observations (non-elites)\\
 \hline
Belonging to upper caste & 0.825 & 0.528 & 1\% & 40 & 3015\\
 Landholding & 0.842 & 0.348 & 1\% & 38 & 3444\\
 Wealth index & 2.575 & 1.01 &  1\% & 40 & 3444 \\
 Stable occupation& 0.525  & 0.067 & 1\% & 40 & 3444\\
 Royal or landlord heritage & 0.3  & 0.009 & 1\% & 40 & 3339\\

 \hline

\end{tabular}

\vskip4em

\begin{tabular}{|p{3cm}||p{3cm}|p{3cm}|p{3cm}| }
 \hline
 \multicolumn{4}{|c|}{Table 4: Structure of occupation in elite HHs by age and gender} \\
 \hline
 Category & Into farming & Into business/salaried job or unemployed with high skill &  Into manual labour or unemployed with low skill\\
 \hline
Male aged 18-40; not studying & 0.59 & 0.35 & 0.06\\
Female aged 18-40; not studying & 0.08 & 0.28 & 0.63\\
All aged 18-40; not studying & 0.35 & 0.33 & 0.31\\
Male aged above 40; not studying & 0.53 & 0.43 & 0.05\\
Female aged above 40; not studying & 0.10 & 0.05 & 0.85\\
All aged above 40; not studying & 0.32 & 0.29 & 0.40\\
\hline

\end{tabular}

\newpage

\begin{tabular}{|p{6cm}||p{6cm}|}
 \hline
 \multicolumn{2}{|c|}{Table 5: Pattern of patronage service provision (from the Elite survey)} \\
 \hline
 Type of services & \% of elite HHs providing\\
 \hline
At least one item \newline from each of E,P and S & 61.10\\
\hline
At least one item \newline from E and P but not S & 7.69\\
\hline
At least one item \newline from E and S but not P & 5.13\\
\hline
At least one item \newline from P and S but not E & 2.56\\
\hline
Only E & 17.95\\
\hline
Only P or only S & 0\\
\hline
\end{tabular}

\newpage

\begin{center}
\begin{figure}[htbp]
\begin{center}
\includegraphics[angle=0, scale=0.90]{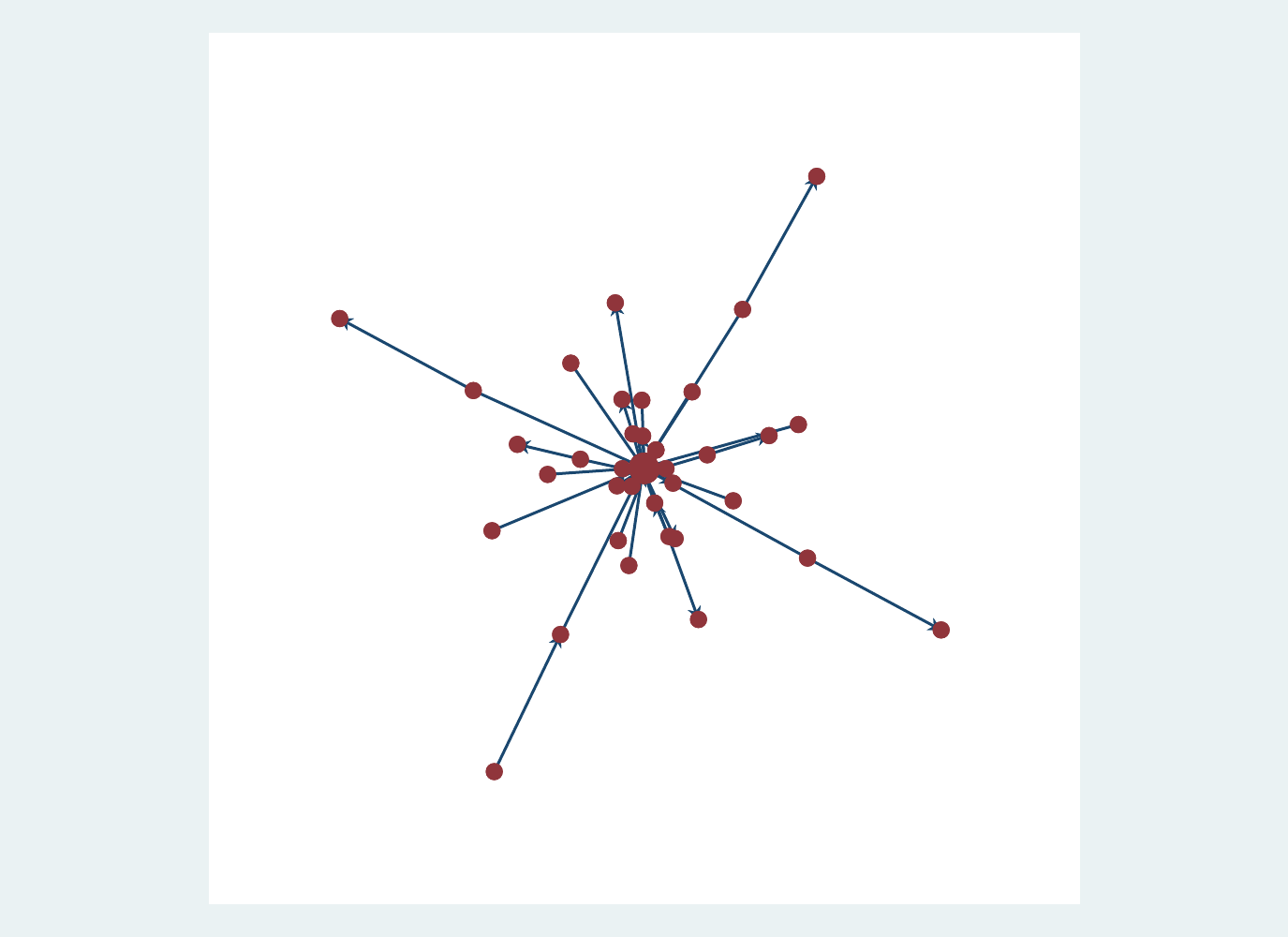}
\caption{Dependence-connections of one elite village}
\label{default}
\end{center}
\end{figure}
\end{center}

\newpage

\begin{center}
\begin{figure}[htbp]
\begin{center}
\includegraphics[angle=0, scale=0.90]{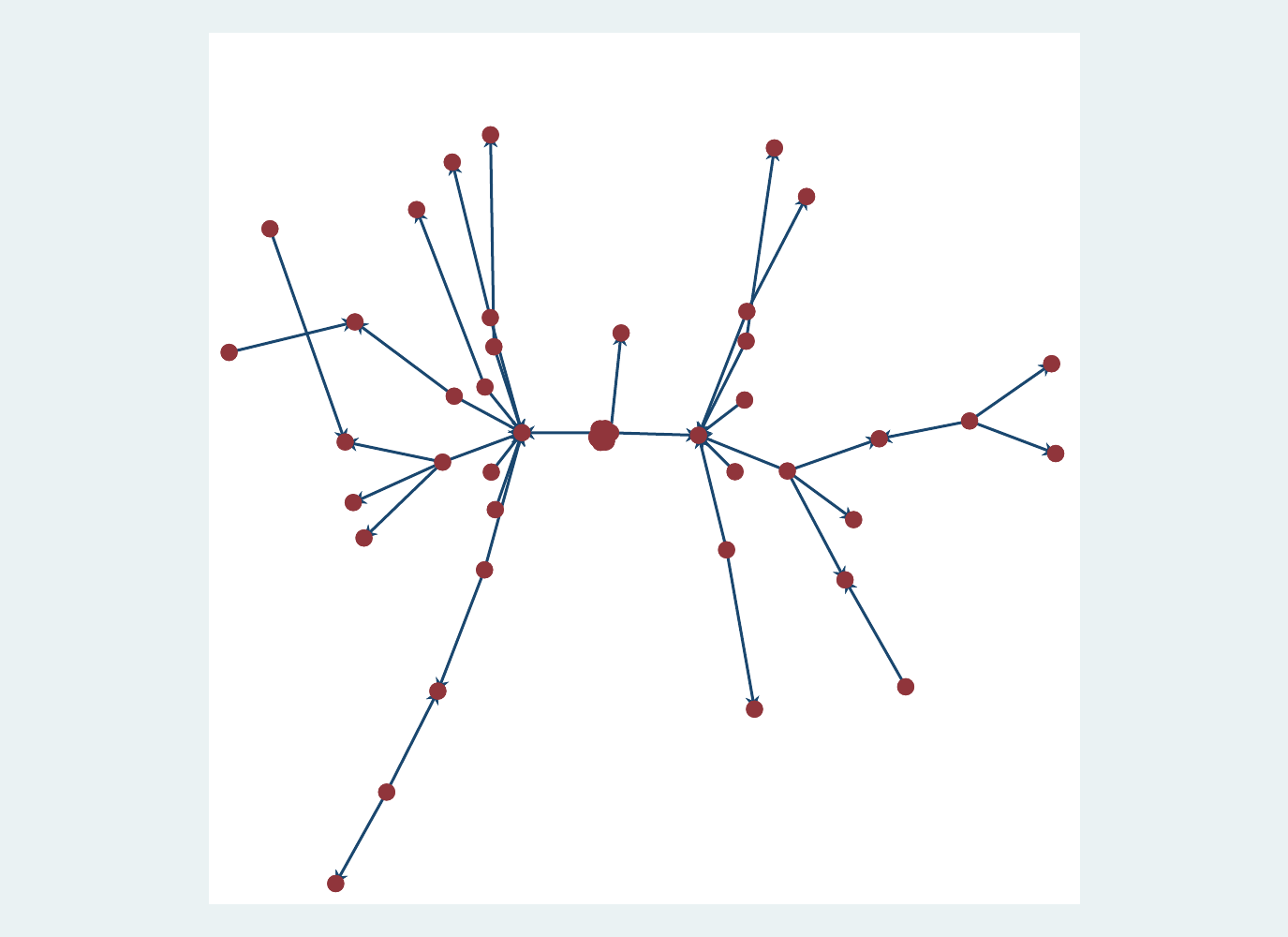}
\caption{Dependence-connections of one non-elite village}
\label{default}
\end{center}
\end{figure}
\end{center}

\newpage

\section*{Appendix: Sample design for LIREP}

Our survey (``Local Institutions and Rural Economic Performances" (LIREP)) sample has a multi-stage, clustered and stratified design. The target sample size was 3600 households based on cost restrictions. As mentioned above, one of the key information that this survey aimed to collect was the local dependence structure and so it was essential to collect information from all or a sizeable number of households in each village. So,  As a result it was decided to select and interview approximately 100 households from each of the selected villages which meant that 36 villages could be selected in the sample.

India is a vast country with 29 states and 7 union territories and each of these regions are culturally and politically different with many policies being implemented at the regional level. To be able to control for these state level effects it was decided to confine the sample to three states (with 12 villages from each state) so that we had sufficient sample sizes at the state level. The three states chosen were Odisha, Maharashtra and (the Eastern part) of Uttar Pradesh (UP). These three states or sub-state regions were chosen for the following reasons. First, these states are located in three different parts of India: Maharashtra in the west, Odisha in the east and UP in the northern part. Secondly, each of these is a major state in India with each having a major different predominant language-based ethnic group. Third, these states have experienced several different historical types of administrative and land-revenue systems during the colonial period (permanent settlement, princely states, taluqdari systems, ryotwari system) which have been shown to have affected the development of post-colonial institutions, and in turn, current economic outcomes. Finally, measuring the institutional impacts of left-wing extremist (LWE hereafter) politics on rural household outcomes was one of our research goals. So, another reason for choosing these states is that LWE activities are prevalent in some parts of each of these states.\\

\noindent {\em Stage 1: Selecting blocks using a stratified design}\\

\noindent Our main research goal was to measure the effect of local level dependence structure on rural household and village level outcomes. But as these local level dependence structures are known to vary by a number of regional factors, we decided to increase the variability of the sample in terms of the local level dependence structures by stratifying the sample by these factors. To increase the variability of the sample along a number of characteristics and to ensure enough sample sizes for one of the variables of interest, left wing extremism, it was decided to stratify the sample along the characteristics as specified below. Most of the information about these stratification variables were available either at the district or at the block (a smaller geographical unit than the district) level. So, it was decided to first select blocks from each of the different strata using probability proportion to size (PPS) sampling where size was measured by the number of households in the block (as in 2001 Census of India, the latest that was available to us) and then select a village randomly from the selected blocks again using PPS sampling method where size was measured by the number of households in the village. The characteristics we used for stratification for each state subsample were as follows:
\begin{itemize}
\item Whether the block had experienced left wing extremist activities (L) or not (NL) between the period 2005 to 2010. This was identified using a number of different sources. 
\item Whether the district containing the block was in a coastal (C) or non-coastal region (NC). This was done directly by using maps. Coastal regions were are expected to have occupational diversity while people in more interior regions (non-coastal) are expected to be mainly in agricultural occupation. To be able to identify different types of dependence, not only predominantly agriculture-based dependence links, the sample was also stratified by coastal and non-coastal region.
\item Whether historically the district was under ryotwari (R) or non-ryotwari (NR) system during the colonial rule. This was identified using the classification provided by Banerjee and Iyer (2005). These historical administrative and land-revenue systems are known to affect post-colonial institutions.
\end{itemize}

Every state did not include all 8 combinations, so we ended up with 13 mutually exclusive and exhaustive strata within the three states (see the Table below). with the added constraint that 12 blocks would have to be selected from each region. As some analysis would look at the LWE impact we also decided that there should be a sufficient number of villages from the LWE stratum. We also needed to select 12 blocks from each state. Our resulting stratification strategy is described in the Table below.

\begin{tabular}{cccc}

\hline 

Stratum No.&State&Stratum&Number of blocks\\

\hline

1&	Eastern UP&	L,NC,NR&	4\\
2&	Eastern UP&	NL,NC,NR&	8\\
3&	Odisha&	L,CO,NR&	2\\
4&	Odisha&	L,NC,NR&	3\\
5&	Odisha&	L,NC,RY&	1\\
6&	Odisha&	NL,CO,NR&	2\\
7&	Odisha&	NL,NC,NR&	3\\
8&	Odisha&	NL,NC,RY&	1\\
9&	Maharashtra&	L,NC,NR&	4\\
10&	Maharashtra&	NL,CO,NR&	1\\
11&	Maharashtra&	NL,CO,RY&	2\\
12&	Maharashtra&	NL,NC,NR&	2\\
13&	Maharashtra&	NL,NC,RY&	3

\end{tabular}

\vskip2em

\noindent {\em Stage 2: Assigning selected blocks to forest and non-forest sub-samples}\\

\noindent At the next sampling stage was we selected one village from each selected block. In the first sampling stage one of the variables we had stratified by was LWE activity. But as blocks are large areas with on average 170 villages (and 50\% of blocks have more than 150 villages but 99\% of blocks have less than 550 villages), not all villages in an LWE affected block will be affected by LWE activity. As it was extremely difficult to get precise information on exactly which of the several hundreds of villages in a block has a history of LWE activities, we decided to indirectly screen for LWE affected villages by selecting villages in these LWE affected blocks that were very near to forest. We used this strategy because forest cover has been found to be highly correlated with LWE activity at least at the district level and there is anecdotal evidence that LWE organisations mainly base their activities in dense forests as state forces find it difficult to enter these areas. We decided to draw two sub-samples from these LWE affected blocks - one from areas next to forests and the other from areas away from forests. We did this by collecting maps of forest cover from the Geological Survey of India and the Forest Research Institute and then overlaying thoese on the maps of villages. We decided to assign the following number of blocks to the forest and non-forest sub-samples of the LWE based strata.

\begin{itemize}

\item Eastern UP- L,NC,NR stratum:: As one of the 4 selected blocks in this strata no forested village, this block was automatically assigned to the Non-forested Sub-Sample and the remaining blocks in that strata, since they summed up to the assigned number of blocks for the Forest Sub-Sample, were allocated to the Forest Sub-Sample.
\item Odisha - L,CO,NR stratum: As one of the selected blocks had no forested village, this block was automatically assigned to the Non-forested Sub-Sample and the remaining blocks in that strata, since they summed up to the assigned number of blocks for the Forest Sub-Sample, were allocated to the Forest Sub-Sample. 
\item Odisha - L,NC,NR stratum: We selected 2 out of the 3 blocks by PPS, where size measure was the proportion of households in forested villages in these blocks, for the Forest Sub-Sample.
\item Odisha - L,NC,RY stratum: The only selected block from this stratum was automatically assigned to the Forest Sub-Sample.
\item Maharashtra - L,NC,NR strata: We selected 3 out of the 4 blocks by PPS where size measure was the proportion of households in forested villages in these blocks, for the Forest Sub-Sample.

\end{itemize}

\noindent {\em Stage 3: Selecting villages from selected blocks}\\

\noindent Finally we selected one village from each of the 36 selected blocks using PPS where size is measured by the total number of households in the village. In two of the villages selected in Maharashtra, battle between the Indian Security Forces and the LWE militants raged  when the HH household survey was to start (in March-April, 2013). So, we had to replace these two villages by two other (safer) villages in the same or adjacent districts with similar administrative and land-revenue history and similar geographical features.\\ 

\noindent {\em Stage 4: Selecting households from selected villages}\\ 

\noindent In villages where the total number of households was less than 100, all households were selected for survey. In each village where the total number of households was more than 100, upto 110 households were selected using simple random sampling. The sampling frame used was the most recent electoral roll for those villages. The target was to interview at least 100 households in each village and at most 110 households. In cases, where due to attrition, non-response etc with the initially chosen sample of HHswe could not reach the target sample size of 100, additional households were selected from the remaining households in the village again using simple random sampling to reach the target sample size. This was repeated until the target sample size was reached. In the final sample, 21 of the sampled villages included less than 50\% of the HHs in the villages, 5 included 50-60\% of the HHs in the villages, 3 included 60-70\% of the HHs in the villages, 2 included 80-95\% of the HHs in the villages and 4 were village censuses.

\end{document}